\documentclass[12pt]{article}

\pdfoutput=1

\usepackage{graphicx}
\usepackage{epsfig}
\usepackage{amsfonts}
\usepackage{amssymb}
\usepackage{amsmath}
\usepackage{dsfont}
\usepackage{bbm}
\usepackage{multirow}
\usepackage[vcentermath]{youngtab}
\usepackage{latexsym}
\usepackage{verbatim}
\usepackage{mcite}
\usepackage{units}
\usepackage{cancel}

\def\beqa{\begin{eqnarray}}
\def\eeqa{\end{eqnarray}}

\def\a{{\alpha}}

\def\d{{\delta}}
\def\m{{\mu}}

\def\bfone{\relax{\rm 1\kern-.35em 1}}

\newcommand{\cM}{{\cal M}}
\newcommand{\cN}{{\cal N}}

\newcommand{\be}{\begin{equation}}
\newcommand{\ee}{\end{equation}}
\newcommand{\ben}{\begin{displaymath}}
\newcommand{\een}{\end{displaymath}}
\newcommand{\bea}{\begin{eqnarray}}
\newcommand{\eea}{\end{eqnarray}}
\newcommand{\nn}{\nonumber}

\newcommand{\bean}{\begin{eqnarray*}}
\newcommand{\eean}{\end{eqnarray*}}

\begin{document}
\pagestyle{plain}

%----------------------------------------------------------------------%
%  numbering sections, equations, footnotes, etc...
%----------------------------------------------------------------------%

\makeatletter \@addtoreset{equation}{section} \makeatother
\renewcommand{\thesection}{\arabic{section}}
\renewcommand{\theequation}{\thesection.\arabic{equation}}
\renewcommand{\thefootnote}{\arabic{footnote}}

%----------------------------------------------------------------------%
%  Resetting of counters
%----------------------------------------------------------------------%

\setcounter{page}{1} \setcounter{footnote}{0}

%----------------------------------------------------------------------%
%  title page
%----------------------------------------------------------------------%

\begin{titlepage}
\begin{flushright}
\small ~~
\end{flushright}

\bigskip

\begin{center}

\vskip 1cm

{\LARGE \bf Critical points of maximal $D=8$ \\[5mm]gauged supergravities} \\[6mm]

\vskip 0.5cm
{\bf Mees de Roo,\, Giuseppe Dibitetto\, and\, Yihao Yin}\\

\vskip 15pt

{\em Centre for Theoretical Physics,\\
University of Groningen, \\
Nijenborgh 4, 9747 AG Groningen, The Netherlands\\
{\small {\tt \{m.de.roo, g.dibitetto, y.yin\}@rug.nl}}} \\

\vskip 0.8cm

\end{center}

\vskip 1cm

\begin{center}

{\bf ABSTRACT}\\[3ex]

\begin{minipage}{13cm}
\small

We study the general deformations of maximal eight-dimensional
supergravity by using the embedding tensor approach. The scalar
potential induced by these gaugings is determined. Subsequently, by
combining duality covariance arguments and algebraic geometry
techniques, we find the complete set of critical points of the
scalar potential. Remarkably, up to SO($2$)$\,\times\,$SO($3$)
rotations there turns out to be a unique theory admitting extrema.
The gauge group of the theory is CSO($2,0,1$).
\end{minipage}

\end{center}

\vfill

\end{titlepage}

%%%%%%%%%%%%%%%%%%%%%%%%%%%%%%%%%%%%%%%%%%%%%%%%%%%%%%%%%
%%
%%               Contents
%%
%%%%%%%%%%%%%%%%%%%%%%%%%%%%%%%%%%%%%%%%%%%%%%%%%%%%%%%%%

\tableofcontents

\section{Introduction}

In the last decade a new formalism has been constructed in extended
supergravity theories which is able to comprise all the consistent
gaugings of a theory in a single universal formulation
\cite{deWit:2002vt, deWit:2004nw, Samtleben:2005bp, deWit:2005hv,
Weidner:2006rp}. This goes under the name of embedding tensor
formalism and it describes deformations of extended supergravities
in a duality covariant way. Indeed the duality group E$_{d(d)}$ of
the ungauged theory in $D=11-d$ dimensions obtained from the
compactification of eleven-dimensional supergravity on a $d-$torus
turns out to determine all the possible deformations (gaugings)
thereof.

The theories in our interest are maximal gauged supergravities in
$D=8$. These theories present, in analogy with half-maximal
supergravity in $D=4$, an SL($2$) factor in the global symmetry
group which allows for gaugings at angles, i.e., gaugings in which
the gauge generators point in different SL($2$) directions. This
feature seems to play the role of the so-called duality angles
\cite{deRoo:1985jh} in half-maximal supergravity in $D=4$, even
though the interpretation of this SL($2$) symmetry as
electromagnetic duality is different in $D=8$, since now the 3-forms
rather than the vectors build SL($2$) doublets with their Hodge
duals.

Some interesting gaugings in $D=8$ were already studied in the
literature, e.g., the SO($3$) gauging found in
ref.~\cite{Salam:1984ft} resulting from compactifying
eleven-dimensional supergravity on $S^3$; furthermore, all the
gaugings in $D=8$ without non-trivial SL($2$) phases have been
classified in terms of their eleven-dimensional origin
\cite{AlonsoAlberca:2003jq, Bergshoeff:2003ri, Roest:2004pk} by
means of a compactification on a group manifold of dimension 3. They
are divided into two categories: in the first one we find all the
gaugings of the type CSO($p,q,r$) \cite{deRoo:2006ms} with
$p+q+r=3$, which arise from a compactification on class A group
manifolds according to the Bianchi classification; in the second
one, we find a set of gaugings which are peculiar because of their
lack of an action principle formulation (class B group manifold
reductions). These theories might stem from the procedure of gauging
the so-called trombone symmetry (see e.g.,
ref.~\cite{LeDiffon:2011wt} where this has been investigated in the
maximal $D=4$ case).

In the context of the embedding tensor, one gives a complete duality
covariant classification of all the gaugings; a natural question to
address is then which of those gaugings actually have a
well-understood eleven-dimensional origin. In contrast with the case
of maximal gauged supergravity theories in $D=9$, where all the
consistent deformations turn out to come from higher dimensions
\cite{FernandezMelgarejo:2011wx}, in $D=8$ there are gaugings for
which no higher-dimensional origin is known yet, e.g., gaugings at
angles.

The main goal of the paper is to derive the scalar potential for the
most general gauging in $D=8$ compatible with maximal supersymmetry
and to study the set of its critical points. The paper is organised
as follows. In section 2 we briefly present the embedding tensor
formalism in $D=8$, we describe the general deformation by means of
group theory and we give the quadratic constraints. In section 3 we
observe that any consistent gauging in $D=8$ is mapped into a
consistent gauging in $D=7$  upon reduction over an $S^1$ and
subsequently we derive the scalar potential of the eight-dimensional
theory by making use of the scalar potential of the
seven-dimensional theory studied in ref.~\cite{Samtleben:2005bp}.
Finally, in section 4, we make use of some algebraic geometry
techniques in order to study the complete landscape of vacua that
these theories have. The main result of this paper is that there is
a unique SO($2$)$\,\times\,$SO($3$) orbit of gaugings of maximal
$D=8$ supergravity allowing for critical points of the scalar
potential. Each of these corresponds to a CSO($2,0,1$) gauging
admitting a non-supersymmetric Minkowski extremum.

\section{Overview on maximal $D=8$ supergavities}

\subsection{The ungauged theory}

The maximal (ungauged) supergravity in $D=8$ can be obtained by
reducing eleven-dimensional supergravity on a $T^3$. The global
symmetry group of this theory is $G_{0}=$
SL($2$)$\,\times\,$SL($3$). The full field content consists of the
following objects (which arrange themselves into irrep's of
$G_{0}$):
\be \textrm{8D :}\qquad \underbrace{e_{\mu}^{\phantom{\mu}a}\,,\,A_{\mu}{}^{Im}\,,\,B_{\mu\nu m}\,,\,C_{\mu\nu\rho}\,,\,L_{m}^{\phantom{m}i}\,,\,\phi\,,\,\chi}_{\textrm{bosonic dof's}}\,\,\,;\,\underbrace{\psi_\mu\,,\,\chi_i\,,}_{\textrm{fermionic dof's}} \ee%
where $\mu,\nu,\cdots$ denote eight-dimensional curved spacetime ,
$a,b,\cdots$ eight-dimensional flat spacetime, $m,n,\cdots$
fundamental SL($3$), $i,j,\cdots$ fundamental SO($3$) and
$I,J,\cdots$ fundamental SL($2$) indices respectively.

The bosonic sector consists of the eight-dimensional vielbein
$e_{\mu}^{\phantom{\mu}a}$, a set of vector fields $A_{\mu}{}^{Im}$,
an SL($3$) triplet of two-forms $B_{\mu\nu m}$, a three-form
$C_{\mu\nu\rho}$ and the scalars $L_{m}^{\phantom{m}i}$ and
$(\phi\,,\,\chi)$ spanning the cosets
$\textrm{SL}(3)/\textrm{SO}(3)$ and $\textrm{SL}(2)/\textrm{SO}(2)$
respectively. The fermionic sector, instead, is made out of a
doublet of symplectic-Majorana (SM) gravitini $\psi_\mu$ and a set
of dilatini $\chi_i$. \newline Let us introduce the following
parametrisations in the scalar sector
\begin{equation}\label{scalarparamWM}
W_{IJ} =\left(
\begin{array}
{cc} e^{-\phi }+\chi ^{2}e^{\phi } & \chi e^{\phi } \\
\chi e^{\phi } & e^{\phi }
\end{array}
\right) \text{ ,}\qquad
M_{mn}=L_{m}^{\phantom{m}i}L_{n}^{\phantom{m}j}\delta_{ij}\,.
\end{equation}
The gravity/scalar part of the action reads
\cite{AlonsoAlberca:2003jq}
\be S=\frac{1}{16\pi G_8}\int
d^8x\,\,e\left(R+\frac{1}{4}\textrm{Tr}(\partial M\partial
M^{-1})+\frac{1}{4}\textrm{Tr}(\partial W\partial
W^{-1})\right)\,,\label{action} \ee
where $e$ is the determinant of the vielbein. The full bosonic
action, in addition to the terms in \eqref{action}, contains kinetic
terms for the vector fields, the two- and three-forms and finally
Chern-Simons terms.

\subsection{Embedding tensor deformations}

When gauging a subgroup of the global symmetry group, the embedding
tensor is turned on, via which in the gauge-covariant derivative the
vectors become coupled to group generators. The embedding tensor
parameterizes the most general deformations consistent with the
global symmetries and supersymmetry. It is an object of the form
$\Theta_{v}^{\phantom{t}\a}$, where the indices $v$ and $\a$ live in
the dual of the representation of the vectors and in the adjoint
representation of the global symmetry group, respectively.

In the maximal $D=8$ case, there
are six vector fields $A_{\mu}{}^{Im}$ transforming in $V'=\left(
\textbf{2},\textbf{3'}\right) $, the dual of
the fundamental representation of $G_{0}$. And there are eleven
group generators, which can be expressed in the adjoint
representation $\mathfrak{g}_{0}$ \footnote{ A traceless pair of
SL($2$) indices $_{I}{}^{J}$ lives in its adjoint representation. So
does a traceless pair $_{m}{}^{n}$ for SL($3$).}:
\begin{subequations}
\begin{eqnarray}
\left[t_{I}{}^{J}\right]_{K}{}^{L}&=&\delta_{I}{}^{L}\delta_{K}{}^{J}-\frac{1}{2}\delta _{I}{}^{J}\delta _{K}{}^{L}\text{ ,} \\%
\left[t_{m}{}^{n}\right]_{p}{}^{q}&=&\delta_{m}{}^{q}\delta_{p}{}^{n}-\frac{1}{3}\delta_{m}{}^{n}\delta_{p}{}^{q}\text{.}%
\end{eqnarray}
\end{subequations}
The embedding tensor $\Theta $ then lives in the representation
$\mathfrak{g} _{0}\otimes V$, which can be decomposed into
irreducible representations as
\begin{equation}
\mathfrak{g}_{0}\otimes V=2\,\cdot\left(
\textbf{2},\textbf{3}\right) \oplus \left(
\textbf{2},\textbf{6'}\right) \oplus  \left(
\textbf{2},\textbf{15}\right) \oplus \left(
\textbf{4},\textbf{3}\right)\text{ .}
\end{equation}
Consistency and supersymmetry restrict the embedding
tensor to the $\left( \textbf{2},\textbf{3}\right) \oplus \left( \textbf{2},\textbf{6'}\right) $ \cite{Weidner:2006rp}. This restriction
goes under the name of linear constraint. It is worth noticing that there are
two copies of the $\left( \textbf{2},\textbf{3} \right) $ irrep in
the above composition; the linear constraint imposes a relation
between them \cite{Samtleben:2008pe}. This shows that, for
consistency, gauging some SL($2$) generators implies the necessity
of gauging some SL($3$) generators as well. Let us denote the
allowed embedding tensor irrep's by $\xi _{Im}$ and $f_{I}{}^{(mn)}$
respectively. Then the following parametrisation holds
\begin{subequations}
\begin{eqnarray}
\Theta_{Im,J}{}^{K}&=&\delta_{I}{}^{K}\xi_{Jm}-\frac{1}{2}\delta_{J}{}^{K}\xi _{Im}\text{ ,} \\%
\Theta_{Im,n}{}^{p}&=&\epsilon_{mnq}f_{I}{}^{qp}-\frac{3}{4}\left(\delta_{m}{}^{p}\xi_{In}-\frac{1}{3}\delta_{n}{}^{p}\xi_{Im}\right) \text{ .}%
\end{eqnarray}
\end{subequations}
Furthermore, one can prove that the generators of the gauge group
can be expressed in the same way:
\begin{subequations}
\label{8D gauge generators}
\begin{eqnarray}
\left(X_{Im}\right)_{J}{}^{K}&=&\Theta_{Im,J^{\prime}}{}^{K^{\prime}}\left[t_{K^{\prime}}{}^{J^{\prime}}\right]_{J}{}^{K}%
=\delta_{I}{}^{K}\xi_{Jm}-\frac{1}{2}\delta_{J}{}^{K}\xi _{Im}\text{ ,} \\%
\left(X_{Im}\right)_{n}{}^{p}&=&\Theta_{Im,n^{\prime}}{}^{p^{\prime}}\left[t_{p^{\prime}}{}^{n^{\prime}}\right]_{n}{}^{p}%
=\epsilon_{mnq}f_{I}{}^{qp}-\frac{3}{4}\left(\delta_{m}{}^{p}\xi_{In}-\frac{1}{3}\delta_{n}{}^{p}\xi_{Im}\right) \text{ .}%
\end{eqnarray}
\end{subequations}
For closure of the algebra, the following quadratic constraints
\cite{Dani:2008} should be imposed on the embedding tensor:
\begin{subequations}
\label{quadratic constraints in 8D}
\begin{eqnarray}
\epsilon ^{IJ}\,\xi _{I p}\xi _{J q} &=&0 \text{ ,}\qquad\qquad
\left( \textbf{1},\textbf{3'}\right)\\%
f_{(I}{}^{np}\xi _{J)p} &=&0 \text{ ,} \qquad\qquad \left(
\textbf{3},\textbf{3'}\right)\\%
\epsilon ^{IJ}\left(\epsilon
_{mqr}f_{I}{}^{qn}f_{J}{}^{rp}+f_{I}{}^{np}\xi _{Jm}\right)
&=&0\text{ .} \qquad\left( \textbf{1},\textbf{3'}\right)\oplus\left( \textbf{1},\textbf{15}\right) %
\end{eqnarray}
\end{subequations}

In this paper, we are mostly interested in the scalar potential in
the Lagrangian, which is quadratic in the embedding tensor. One can
write down an Ansatz for such a potential\footnote{The corresponding
term to be added to the Lagrangian \eqref{action} should be
$\mathcal{L}_{V}=-eg^{2}V$ , where $g$ is the coupling strength.}:
\begin{equation}
V=W^{IJ}\,\left[f_{I}{}^{mn}f_{J}{}^{pq}\left(a\,M_{mp}M_{nq}+b\,M_{mn}M_{pq}\right)+c\,\xi_{Im}\xi_{Jn}M^{mn}\right] \text{ ,}%
\label{Ansatz 8D V}%
\end{equation}
where $W_{IJ}$ and $M_{mn}$ are elements of the scalar cosets
introduced in \eqref{scalarparamWM}, whereas  $W^{IJ}$ and $M^{mn}$
denote their inverse matrices and $a$, $b$ and $c$ are coefficients
that are going to be determined. The most convenient way of fixing
these coefficients is to use the scalar potential in maximal $D=7$
supergravity, which was already well studied in
\cite{Samtleben:2005bp}.

\section{Gaugings of $D=8$ supergravity as truncations of gaugings in $D=7$ }
\subsection{Review of maximal $D=7$ supergravity}
The general deformations of seven-dimensional maximal supergravity
are constructed and presented in ref.~\cite{Samtleben:2005bp}. For
the sake of clarity, we briefly summarise the results obtained
there. The global symmetry group is SL($5$), which has an adjoint
representation \textbf{24}. The vectors\footnote{Here we denote by
$M$ a fundamental SL($5$) index.} $A_{\m}{}^{MN}=A_{\m}{}^{[MN]}$ of
the theory transform in the \textbf{10'} of SL($5$). Then the
embedding tensor $\Theta$ will take values in the following irrep's
of SL($5$)
\be
\textbf{10}\otimes\textbf{24}\,=\,\textbf{10}\oplus\textbf{15}\oplus\textbf{40'}\oplus
\textbf{175}\,.\ee
After imposing the linear constraint, the parametrization of the embedding tensor
is restricted to only two irreducible components $\textbf{15}\,\oplus\, \textbf{40'}$:
\begin{subequations}
\bea Y_{MN}&=&Y_{(MN)}\phantom{aaaaaaaaaaaaaaaaaaaaa}
\textbf{15}\,:\phantom{aaaaaaaaaaa}\,\yng(2)\,\,,\\[4mm]
 Z^{MN,P}&=&Z^{[MN],P}\textrm{  with
}Z^{[MN,P]}=0\qquad\,\textbf{40'}\,:\,\yng(1,1)\,\,\otimes\,\,\yng(1)=\xcancel{\hspace{3mm}\yng(1,1,1)\hspace{3mm}}\,\,\oplus\,\,\yng(2,1)\,\,,\label{linZ}
\eea
\end{subequations}
where $M,N$ and $P$ represent fundamental SL($5$) indices.
Furthermore, supersymmetry and the consistency of the gauging
require the following quadratic constraints to hold
\begin{equation}
Y_{MQ}\,Z^{QN,P}+2\,\epsilon_{MRSTU}\,Z^{RS,N}\,Z^{TU,P}\,=\,0\text{
.} \label{7D QC2}
\end{equation}
Any embedding tensor configuration satisfying \eqref{7D QC2}
identifies a gauging of a certain (at most) ten-dimensional group
suitably embedded in SL($5$). The expression of the gauge generators
is given by \cite{Samtleben:2005bp}
\be\label{7D X}
\left(X_{MN}\right)_P^{\phantom{P}Q}=\d_{[M}^{\phantom{[M}Q}Y_{N]P}-2\,\epsilon_{MNPRS}Z^{RS,Q}\,,
\ee
where the pair of indices $_P^{\phantom{P}Q}\,$ is in the adjoint
representation of SL($5$) once the linear constraint is satisfied.

The scalar sector is described by the SL($5$)$/$SO($5$) coset
geometry parametrised by the symmetric matrix $\cM_{MN}$ with
inverse $\cM^{MN}$. This divides the isometry group of the scalar
manifold SL($5$) into unphysical scalar degrees of freedom
(generating the adjoint representation of SO($5$)) and physical
scalar fields completing them to the $\textbf{24}$, i.e., the
adjoint representation of SL($5$). Maximal supersymmetry completely
and uniquely determines the scalar potential to be of the form
\bea V&=&
\frac{1}{64}\bigg(2\cM^{MN}Y_{NP}\cM^{PQ}Y_{QM}-(\cM^{MN}Y_{MN})^2\bigg)+\nn\\
&+&Z^{MN,P}Z^{QR,S}\bigg(\cM_{MQ}\cM_{NR}\cM_{PS}-\cM_{MQ}\cM_{NP}\cM_{RS}\bigg)\,.\label{7D
V} \eea

\subsection{From $D=7$ to $D=8$}
Every gauging in $D=8$ must be an at most six-dimensional subgroup
of the global symmetry group SL($2$)$\,\times\,$SL($3$). After
dimensional reduction to $D=7$, the global symmetry group gets
enhanced with respect to what one would naively expect\footnote{One
would expect
$\mathbb{R}^+\,\times\,\textrm{SL}(2)\,\times\,\textrm{SL}(3)$,
whereas it turns out to be enlarged to an $\textrm{SL}(5)$.}; for
this reason, one would certainly expect any consistent gauging of
the eight-dimensional theory to be reduced to a consistent gauging
of the seven-dimensional theory where the gauge group, though,
undergoes an enlargement just in the same way as for the global
symmetry group. This statement implies that the irreducible
components of the embedding tensor in eight dimensions must be
obtained as a truncation of the embedding tensor in $D=7$. This
implies the possibility of deriving the scalar potential of maximal
$D=8$ gauged supergravity from the expression of the
seven-dimensional scalar potential given in \eqref{7D V}, after
understanding how the eight-dimensional degrees of freedom
associated with internal symmetries sit inside SL($5$) irrep's. To
this end, we need the branching of some relevant irrep's of SL($5$)
with respect to irrep's of SL($2$)$\,\times\,$SL($3$), which is a
maximal subgroup thereof. The embedding turns out to be unique and
it gives rise to the following decompositions
\begin{subequations}
\bea \textbf{5}\quad &\longrightarrow &\quad
(\textbf{2},\textbf{1})\,\oplus\,(\textbf{1},\textbf{3})\,,\label{br5}\\
\textbf{15}\quad &\longrightarrow &\quad
(\textbf{1},\textbf{6})\,\oplus\,(\textbf{2},\textbf{3})\,\oplus\,(\textbf{3},\textbf{1})\,,\label{br15}\\
\textbf{24}\quad &\longrightarrow &\quad
(\textbf{1},\textbf{1})\,\oplus\,(\textbf{1},\textbf{8})\,\oplus\,(\textbf{2},\textbf{3})\,\oplus\,(\textbf{2},\textbf{3'})\,\oplus\,(\textbf{3},\textbf{1})\,,\label{br24}\\
\textbf{40'}\quad &\longrightarrow &\quad
(\textbf{1},\textbf{3'})\,\oplus\,(\textbf{1},\textbf{8})\,\oplus\,(\textbf{2},\textbf{1})\,\oplus\,(\textbf{2},\textbf{6'})\,\oplus\,(\textbf{2},\textbf{3})\,\oplus\,(\textbf{3},\textbf{3'})\,.\label{br40p}\eea
\end{subequations}
The decomposition \eqref{br5} essentially tells that the fundamental
SL($5$) index $M=1,2,3,4,5$ goes into $(I\,;\,m)$, where $I=+,-$ and
$m=1,2,3$ represent fundamental SL($2$) and SL($3$) indices
respectively. The decomposition \eqref{br24} tells us how the
SL($2$)$\,\times\,$SL($3$) scalar degrees of freedom (living in the
$(\textbf{1},\textbf{8})\,\oplus\,(\textbf{3},\textbf{1})$) are
embedded in the adjoint of SL($5$). It is worth mentioning at this
point that we are losing a Cartan generator in the branching
procedure; such an abelian generator is realised as an extra
$\mathbb{R}^+$ factor corresponding to a dilaton in the
seven-dimensional theory, with respect to which any
eight-dimensional object should have a scaling weight which we are
omitting. This extra scalar exactly accounts for the
$(\textbf{1},\textbf{1})$ irrep appearing in \eqref{br24}. The
truncation that we need consists then in switching off all the
off-diagonal axionic excitations (spanning the
$(\textbf{2},\textbf{3})$ and $(\textbf{2},\textbf{3'})$ terms in
\eqref{br24}), thus resulting in the following parametrisation
\be \cM_{MN}=\left(
\begin{array}{c|c}
e^{3\sigma}\,W_{IJ} & 0 \\[1mm]
\hline
\\[-4mm]
0 & e^{-2\sigma}\,M_{mn}
\end{array}
\right)\,,
\label{scalar coset decomposition}%
\ee
where $\sigma$ is the extra dilaton corresponding to $\mathbb{R}^+$,
whereas $W_{IJ}$ and $M_{mn}$ parametrise the SL($2$)$/$SO($2$) and
SL($3$)$/$SO($3$) cosets respectively. It has been checked
explicitly that the scaling weights of all the terms in the $D=8$
scalar potential with respect to the extra $\mathbb{R}^+$ are all
equal such that it is perfectly consistent to set $\sigma=0$ in the
rest of our derivation, since any other constant value can be seen
as a change of normalisation of the potential energy in the
lagrangian.

As has been mentioned, the embedding tensor in maximal $D=8$
supergravity lives in the \cite{Weidner:2006rp}
$(\textbf{2},\textbf{3})\,\oplus\,(\textbf{2},\textbf{6'})$, which
are parametrised by $\xi_{Im}$ and
$f_I^{\phantom{I}mn}=f_I^{\phantom{I}(mn)}$, respectively. After
taking a look at the decompositions given in \eqref{br15} and in
\eqref{br40p}, one can infer that $\xi$ will in general source
non-vanishing components of both $Y$ and $Z$, whereas $f$ will only
turn on components of\footnote{This is due to the fact that a
$(\textbf{2},\textbf{3})$ irrep appears in the branching of both the
$\textbf{15}$ and the $\textbf{40'}$, whereas a
$(\textbf{2},\textbf{6'})$ is only present in the decomposition of
the $\textbf{40'}$.} $Z$. This results in the following general
Ansatz
\begin{subequations}
\label{Ansatz embedding tensor decomposition from 7D to 8D}
\begin{eqnarray}
Z^{Im,n}=-Z^{mI,n} &=& \lambda_{1}\,\epsilon^{IJ}f_{J}{}^{mn}+\lambda_{2}\,\epsilon^{mnp}\epsilon^{IJ}\xi_{Jp}\text{ ,}\label{Z1}\\%
Z^{mn,I}&=& \lambda_{3}\,\epsilon^{mnp}\epsilon^{IJ}\xi_{Jp}\text{ ,}\label{Z2}\\%
Y_{Im}=Y_{mI}&=& \lambda_{4}\,\xi_{Im}\text{ ,}\label{Y}%
\end{eqnarray}
\end{subequations}
where all the other components of $Y$ and $Z$ vanish and the
parameters $\lambda_1,\lambda_2,\lambda_3,\lambda_4$ will be fixed
by some consistency requirements. First of all, the linear
constraint implies in particular that $Z$ lives in the
$\textbf{40'}$, which means, as explained in \eqref{linZ}, that the
three-form must vanish
\begin{equation}
Z^{[MN,P]}=0\text{ ,}
\end{equation}
which yields the condition
\begin{equation}
\lambda _{3}=-2\,\lambda _{2}\text{ .}  \label{lambda32}
\end{equation}
Secondly, we will substitute the Ansatz (\ref{Ansatz
embedding
tensor decomposition from 7D to 8D}) into (\ref{7D X}), which translates into the following expression for the gauge generators\footnote{We use the convention that $\epsilon _{IJmnp}=\epsilon _{IJ}\epsilon _{mnp}$.}%
\begin{subequations}
\label{X decomposition with lambda}
\begin{eqnarray}  \label{Decomposition of 7D gauge generators}
\left( X_{Im}\right) _{n}{}^{p} &=&4\lambda _{1}\epsilon_{mnq}f_{I}{}^{qp}+\left( 4\lambda _{2}-\frac{1}{2}\lambda_{4}\right) \delta _{m}{}^{p}\xi _{In}-4\lambda _{2}\delta_{n}{}^{p}\xi _{Im}\text{ ,}%
\label{XImnp} \\%
\left( X_{Im}\right) _{J}{}^{K} &=&\left( 4\lambda_{3}+\frac{1}{2}\lambda _{4}\right) \delta _{I}{}^{K}\xi_{Jm}-4\lambda _{3}\delta _{J}{}^{K}\xi_{Im}\text{ ,}%
\label{XImJK} \\%
\left( X_{IJ}\right) _{m}{}^{K} &=&\left( -8\lambda _{3}+\lambda_{4}\right)\delta _{\lbrack I}{}^{K}\xi _{J]m}\text{ ,}%
\label{XIJmK} \\%
\left( X_{mn}\right) _{I}{}^{p} &=&4\lambda _{1}\epsilon_{mnq}f_{I}{}^{qp}+\left(8\lambda _{2}+\lambda _{4}\right)\delta _{\lbrack m}{}^{p}\xi_{|I|n]}\text{ ,}%
\label{XmnIp}%
\end{eqnarray}
\end{subequations}
the remaining components being all zero. Now one has to make sure
that the expression of the eight-dimensional gauge generators given
in (\ref{8D gauge generators}) is correctly obtained.

Therefore, by comparing (\ref{8D gauge generators}) with
(\ref{XImnp}) and (\ref{XImJK})%
\footnote{(\ref{XIJmK}) and (\ref{XmnIp}) are some extra
non-vanishing gauge generators due to the enlargement of the gauge
group we already mentioned when compactifying from $D=8$ to $D=7$.},
while also taking (\ref{lambda32}) into account, one can
consistently fix all $\lambda$'s as:
\begin{equation}
\lambda _{1}=\frac{1}{4},\ \lambda _{2}=-\frac{1}{16},\ \lambda_{3}=\frac{1 }{8},\ \lambda _{4}=1\text{ .}%
\label{lambda values}
\end{equation}
By substituting these values into (\ref{Ansatz embedding tensor
decomposition from 7D to 8D}), the decomposition rules on the
embedding tensor are obtained:
\begin{subequations}
\label{embedding tensor decomposition from 7D to 8D}
\begin{eqnarray}
Z^{Im,n}=-Z^{mI,n} &=&\frac{1}{4}\,\epsilon^{IJ}f_{J}{}^{mn}-\frac{1}{16}\,\epsilon ^{mnp}\epsilon ^{IJ}\xi _{Jp}\text{ ,}  \label{Z1} \\%
Z^{mn,I} &=&\frac{1}{8}\,\epsilon ^{mnp}\epsilon ^{IJ}\xi_{Jp}\text{ ,}  \label{Z2} \\%
Y_{Im}=Y_{mI} &=&\xi _{Im}\text{ ,}  \label{Y}\\%
\text{other components}&=& 0 \text{ .}
\end{eqnarray}

Furthermore, one can check that substituting (\ref{embedding tensor
decomposition from 7D to 8D}) into the $D=7$ quadratic constraints
(\ref{7D QC2}) exactly leads to the ones in $D=8$ as shown in
(\ref{quadratic constraints in 8D}).
\end{subequations}

Finally in this section, let's come back to the scalar potential.
One can apply the decomposition rules (\ref{scalar coset
decomposition}) and (\ref{embedding tensor decomposition from 7D to
8D}) on the $D=7$ scalar potential (\ref{7D V}), so that the
relative coefficients in (\ref{Ansatz 8D V}) can be determined, and
by taking the normalisation of the action \eqref{action} into
account one can further fix the overall factor of (\ref{Ansatz 8D
V}). Then the $D=8$ scalar potential is fully derived:
\begin{equation}
\label{8D V}
V= \frac{1}{2}\,W^{IJ}\,\left[f_{I}{}^{mn}f_{J}{}^{pq}\,\left(2M_{mp}M_{nq}-M_{mn}M_{pq}\right)+\xi_{Im}\xi_{Jn}M^{mn}\right] \text{ .}%
\end{equation}

\section{Investigating the vacua}
\subsection{Extrema of the potential}
In the previous sections we have presented the quadratic constraints
(\ref {quadratic constraints in 8D}) and the scalar potential
(\ref{8D V}). With these formulae in hand we can now investigate the
vacua of the maximal $D=8$ supergravity.

In total there are 7 scalars for the coset
$\frac{\textrm{SL}(2)}{\textrm{SO}(2)}\,\times\,\frac{\textrm{SL}(3)}{\textrm{SO}(3)}$.
In \eqref{scalarparamWM} we already gave a parametrisation for the
SL($2$) scalars; now we also specify a parametrisation of the
vielbein $L$ appearing in \eqref{scalarparamWM} containing the
information about the SL($3$) scalars, which is given by
\be L_{m}^{\phantom{m}i} =\left(
\begin{array}{ccc}
e^{-\phi _{1}} & \chi _{1}e^{\frac{\phi _{1}-\phi _{2}}{2}} & \chi_{2}e^{\frac{\phi _{1}+\phi _{2}}{2}} \\%
0 & e^{\frac{\phi _{1}-\phi _{2}}{2}} & \chi _{3}e^{\frac{\phi _{1}+\phi _{2}}{2}} \\%
0 & 0 & e^{\frac{\phi _{1}+\phi _{2}}{2}}%
\end{array}
\right)
\text{ .}  \label{scalar param. L}%
\ee
Subsequently, by substituting such a parametrisation into the scalar
potential (\ref{8D V}) and requiring that
\begin{equation}
\frac{\delta V}{\delta \text{ (scalars)}}=0\text{ ,}
\end{equation}
one obtains 7 equations which represent the extremality condition
for the scalar potential. Since the full theory enjoys a global
SL($2$)$\,\times\,$SL($3$) duality symmetry, one can choose to solve
these equations in the origin of moduli space (setting all 7 scalars
to zero\footnote{This translates into $W=\mathds{1}_2$ and
$M=\mathds{1}_3$, from which it becomes manifest that the origin
still presents a residual SO($2$)$\,\times\,$SO($3$)
invariance.\label{footnote_origin}}.). This can always be done
without loss of generality by performing a non-compact duality
transformation. This will translate the 7 equations of motion for
the scalars into a set of 7 quadratic conditions in the embedding
tensor components. Furthermore the quadratic constraints (\ref
{quadratic constraints in 8D}) give another 30 equations in the
embedding tensor components which need to be satisfied for the
solution to be consistent. This set of 37 equations appears in the
form an ideal consisting of homogeneous polynomial equations which
can be solved for the components $\xi _{Im}$ and $f_{I}{}^{(mn)}$.

As explained in the footnote \ref{footnote_origin}, we still have
compact duality transformations that we can use in order to simplify
the general form of $\xi$ and $f$ without spoiling the choice of
solving the equations of motion in the origin. For instance, we can
make use of an SO($3$) transformation in order to diagonalise
$f_{-}{}^{mn}$, whereas for the moment we don't need to exploit
SO($2$) transformations.

In the next step, we will exploit an algebraic geometry tool called
the Gianni-Trager-Zacharias (GTZ) algorithm \cite{GTZ}. This
algorithm has been computationally implemented by the
\textsc{\,Singular\,} project \cite{DGPS} and such an implementation
has been used recently in ref.~\cite{Dibitetto:2011gm} for a purpose
similar to the one discussed here.

We find in the end only one SO($2$)$\,\times\,$SO($3$) orbit of
solutions\footnote{It is worth mentioning that an
SO($2$)$\,\times\,$SO($3$) rotation has been used in order to reduce
some apparently inequivalent solutions to the form
\eqref{solution}.}, in which the simplest representative is given by
\be f_{+}{}^{mn}= \left(
\begin{array}{ccc}
\lambda & 0 & 0\\
0 & \lambda & 0\\
0 & 0 & 0
\end{array}
\right)\text{ ,}
\,\,\,\,\,f_{-}{}^{mn}\,\,=\,\,\xi_{+m}\,\,=\,\,\xi_{-m}\,\,=\,\,0\text{
,}\label{solution}\ee
where $\lambda$ represents an arbitrary real parameter. This orbit
of solutions represents a CSO($2,0,1$) gauging, which was obtained
in ref.~\cite{Roest:2004pk} as eleven-dimensional supergravity
compactified on an ISO($2$) manifold, with structure constants given
by $f_{mn}^{\phantom{mn}p}=\epsilon_{mnq}f_{+}{}^{qp}$.

\subsection{Supersymmetry breaking analysis}
Let's now see whether we can say something about the fraction of
supersymmetry preserved by this class of solutions. Using the
expression in ref.~\cite{AlonsoAlberca:2003jq} for the variation of
the gravitino\footnote{This expression is valid for a maximally
symmetric solution of a theory obtained from the reduction of
eleven-dimensional supergravity on a three-dimensional group
manifold with structure constants $f_{mn}{}^{p}$.}
\be \delta
\psi_{\mu}=-\frac{g}{48}e^{\phi/2}\,f_{mn}{}^{p}\Gamma^{mn}{}_{p}\Gamma_{\mu}\,\varepsilon\,,
\ee
and choosing the following parametrisation for eleven-dimensional
Dirac matrices \cite{Salam:1984ft}
\be \Gamma^{\mu}=\gamma^{\mu}\otimes \mathds{1}_2
\qquad\textrm{and}\qquad \Gamma^{m}=\gamma^{9}\otimes \sigma^m\,,\ee
one finds
\be \delta \psi_{\mu}=
-\frac{g}{24}e^{\phi/2}f_{+}{}^{mn}M_{mn}\,\gamma^{9}\gamma_{\mu}\,\varepsilon\,\propto\,2\lambda\,\gamma^{9}\gamma_{\mu}\,\varepsilon\,\neq
0\,,\ee
which implies that these solutions are always non-supersymmetric,
whereas for $\lambda=0$ the standard supersymmetric Minkowski vacuum
of the ungauged theory is recovered.

\subsection{Stability analysis}
To this end, we need to compute the 7 eigenvalues of the mass matrix
at the solution. We would like to point out that the scalars
parametrised in \eqref{scalarparamWM} and \eqref{scalar param. L}
are not canonically normalised, i.e., the kinetic terms read
\bea \mathcal{L}_{\textrm{kin}}&=&\frac{1}{2}\,K_{ij}\left(\partial
\Phi^i\right) \left(\partial \Phi^j \right)\,\,\,\textrm{, } \\
K_{ij}&=&\left(
\begin{array}{ccccccc}
 1 & 0 & 0 & 0 & 0 & 0 & 0 \\
 0 & 3 & 0 & 0 & 0 & 0 & 0 \\
 0 & 0 & 1 & 0 & 0 & 0 & 0 \\
 0 & 0 & 0 & e^{2 \phi } & 0 & 0 & 0 \\
 0 & 0 & 0 & 0 & e^{3 \phi_1-\phi_2 } & 0 & 0 \\
 0 & 0 & 0 & 0 & 0 & e^{3 \phi_1+\phi_2} & -e^{3\phi_1+\phi_2} \chi_1 \\
 0 & 0 & 0 & 0 & 0 & -e^{3\phi_1+\phi_2} \chi_1 & e^{3\phi_1+\phi_2}\chi_1^2+e^{2\phi_2}
\end{array}
\right)\,. \label{Kparam.}\eea
This means that the physical mass matrix is given by
\be \left(m^2\right)^i_{\phantom{i}j}=K^{ik}\,\partial_k\,\partial_j
V\,,\label{masses} \ee
where $K^{ij}$ denotes the inverse of the matrix $K$ in
\eqref{Kparam.}.

Computing the mass matrix defined in \eqref{masses} for the
solutions \eqref{solution}, one finds the following eigenvalues
\be
\begin{array}{cccc}
0\,\,(\times\,5)\,, & & & 8\,\lambda^2\,\,(\times\,2)\,.
\end{array}
\ee
These solutions present five flat directions. In fact, the
underlying CSO gauging doesn't have any non-trivial SL($2$) phases
($f_+\neq0$, and $f_-=0$) and hence the potential given in \eqref{8D
V} has an overall $e^{\phi}$ and no dependence at all on $\chi$.
Only the vanishing of $V$ itself at the solution saves it from the
run-away. This explains why the SL($2$) scalars are massless.

Furthermore, in any theory in which a bosonic symmetry is gauged,
one expects a number of Goldstone bosons corresponding to unbroken
generators of the gauge group. This explains the presence of some
extra flat directions. Discussion of the stability of all flat
directions would require an analysis of higher-order derivatives.

\section{Conclusions}

In the present paper we considered the general deformations of
maximal $D=8$ supergravity and we have derived the scalar potential
for the general case. Subsequently, by combining duality covariance
arguments with algebraic geometry techniques, we were able to study
the set of extremality conditions for the general gauging. The
remarkable outcome is that there is only a unique
SO($2$)$\,\times\,$SO($3$) orbit of Minkowski solutions
corresponding with a CSO($2,0,1$) gauging.  As discussed above, they
are all non-supersymmetric with no possibility of an intermediate
case of partial supersymmetry breaking from $\cN=2$ to $\cN=1$.
Moreover, these solutions have the good feature of being free of
tachyons at a quadratic level. There are, though, a number of flat
directions which might require a further analysis at higher
perturbative orders.

%%%%%%%%%%%%%%%%%%%%%%%%%%%%%%%%%%%%
%
% Acknowledgments
%
%%%%%%%%%%%%%%%%%%%%%%%%%%%%%%%%%%%%

\vspace*{5mm} \noindent {\bf \large Acknowledgments}

\vspace*{3mm} \noindent We are grateful to Dani Puigdom\`enech,
Diederik Roest and Henning Samtleben for very interesting and
stimulating discussions. The work of G.D. is supported by a VIDI
grant from the Netherlands Organisation for Scientific Research
(NWO). The work of Y.Y. is supported by the Ubbo Emmius Programme
administered by the Graduate School of Science, University of
Groningen.

\bibliography{biblio}
\bibliographystyle{utphys}

\end{document}